\documentclass{amsart}
\usepackage{lipsum}% http://ctan.org/pkg/lipsum
\usepackage{amsaddr}
\usepackage{graphicx}        % standard LaTeX graphics tool
\usepackage{float}        % standard LaTeX graphics tool
\begin{document}
\title{Oscillating behavior within the social force model}
\author{Mohcine Chraibi}
\address{J\"ulich Supercomputing Centre, Forschungszentrum J\"ulich  GmbH. 52425 J\"ulich, Germany}
\email{m.chraibi@fz-juelich.de}

\date{\today}

\maketitle

\section{Introduction}

The social force model (SFM) \cite{Helbing2000} belongs to a class of microscopic force-based pedestrian model for which the interaction with the neighbors solely depends on the distance spacings. 
Yet, distance based models can lead to unrealistic oscillating behaviors with collision and negative speed (especially in 1D, see for instance \cite{Treiber2013} and the harmonic oscillator equation).  With oscillation we describe a situation where pedestrians perform, in the direction of the intended movement, forward and backward movement. This behavior is to be distinguished from the oscillations in the passing direction in bottlenecks \cite{Helbing2001b} and from lateral oscillations along the shoulder direction \cite{ Hoogendoorn2005,LiuX2009}. 

The problem of oscillations in the movement direction was investigated in the literature and shown to rise in general in simulations with force-based models \cite{Chraibi2011,Koester2013,Dietrich2014a}. In \cite{Chraibi2010a} the model-induced oscillations were quantified by introducing the  ``oscillation-proportion''  introduced. Furthermore, it was suggested that this phenomena is the dual-problem of another model-induced model, namely the overlapping problem.

In this article we show on basis of analytical investigation of a simplified one-dimensional scenario, that only particular values of the parameters allow collision-free and non-oscillating simulation.  However, `realistic'' values for the (physical) parameters of the SFM do not produce such simulations.

\section{Local stability analysis}

The social force model in one dimension and with a coupled interaction with the predecessor writes:
 \begin{equation}
\ddot x_n =  -a \exp\Big(\frac{-\Delta x_n}{b}\Big) + \frac{v_0-\dot x_n}{\tau},
\label{eq:sfm}
\end{equation}
with $v_0,a,b,\tau>0$. 
Hereby pedestrians are presented as points and $\Delta x_n = x_{n+1} - x_n$. 

Equation (\ref{eq:sfm}) can be linearly approximated for $\Delta x_n / b \approx 0$
as follows
\begin{equation}
\ddot x_n \approx  -a \Big(1+\frac{-\Delta x_n}{b}\Big) + \frac{-\dot
  x_n}{\tau} + \tilde v_0,\;\;\; \tilde v_0=\frac{v_0}{\tau},
\end{equation}
or 
\begin{equation}
\ddot x_n +\frac{\dot x_n}{\tau} -\frac{a}{b}\Delta x_n +  a  - \tilde
v_0\approx 0.
\end{equation}
Using the following variable substitution 
$\xi= -\frac{a}{b}\Delta x_n +  a  - \tilde
v_0,$
%\;\rightarrow \dot \xi = \frac{a}{b} \dot x_n\;\;(x_{n+1}=0)$
and assuming that $x_{n+1}$ is constant we get
\begin{equation}
\frac{b}{a} \ddot \xi + \frac{b}{a\tau}\dot \xi + \xi=0,
\end{equation}
which can be brought in the form of a harmonic oscillator
\begin{equation}
 \ddot \xi + r\dot \xi +\omega_0^2 \xi=0,
\label{eq:1}
\end{equation}
with $r=\frac{1}{\tau}$ and $\omega_0^2=\frac{a}{b}$.

Since the coefficients $r$ and $\omega_0$ are constant, 
(\ref{eq:1}) will have a solution of the form  $\xi = e^{\lambda t}$
which yields the following characteristic polynomial:
\begin{equation}
\lambda^2 + r \lambda + \omega_0^2  = 0.
\label{eq:2}
\end{equation}
The roots of (\ref{eq:2}) are 
\begin{equation}
\lambda_{1,2} = \frac{-r \pm \sqrt{r^2 - 4 \omega_0^2}}{2},
\end{equation}
which gives a general solution of the form 
$$\xi = c_1 e^{\lambda_1} + c_2 e^{\lambda_2}.$$

The system described by (\ref{eq:sfm}) does not oscillate if the imaginary part of the solutions are nil, i.e. if $r^2-4\omega_0^2>0$ or
\begin{equation}
\frac{a}{b}\tau^2 <\frac{1}{4}.
\label{eq:cond1}
\end{equation}
Note that the solution does not depend $explicitly$  on $v_0$.

% the condition for that the solution is oscillating or not does not depend on this parameter. 

\section{Interpretation}

In \cite{Werner2003} is has been reported that evaluation of empirical data yields $\tau = 0.61$ s. A slightly different value of $\tau$ was measured in \cite{Moussaid2009a} ($\tau = 0.54 \pm 0.05$ s). It simulation   $\tau=0.5$ s is often used as a realistic value.

\begin{figure}[H]
\centering
\includegraphics[scale=0.5]{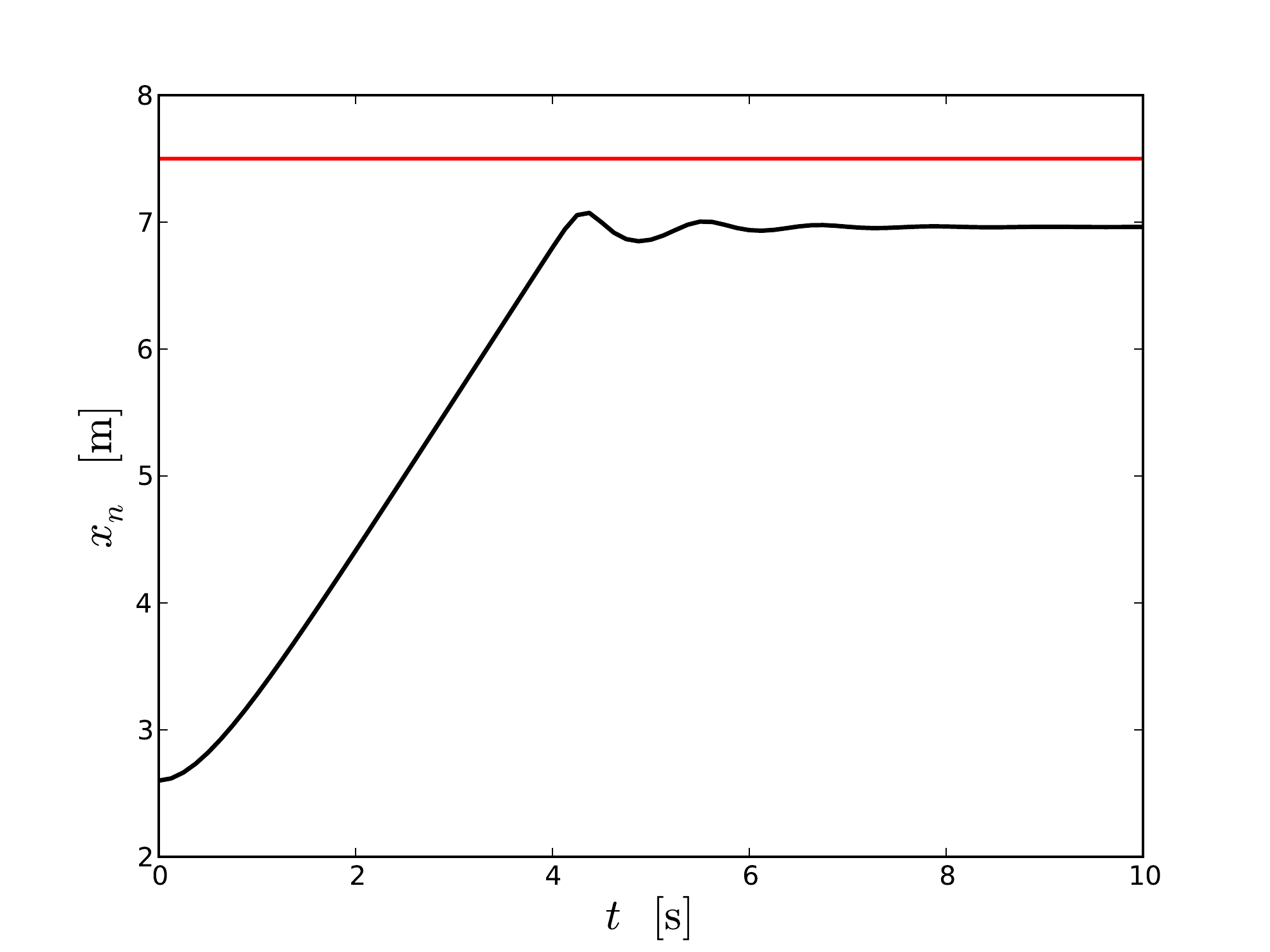}
\caption{Simulation with $a=2000$ N and $b=0.08$ m.}
\label{fig1}
\end{figure}

In the following examples, we always assume that $v_0=1.2$ m/s and $\tau=0.5$ s. 
The parameter values in  \cite{Helbing2000, Parisi2009} $a = 2000$\, N and  $b =
0.08$\, m lead to oscillations. See  Fig. \ref{fig1}

\begin{figure}[H]
\centering
\includegraphics[scale=0.5]{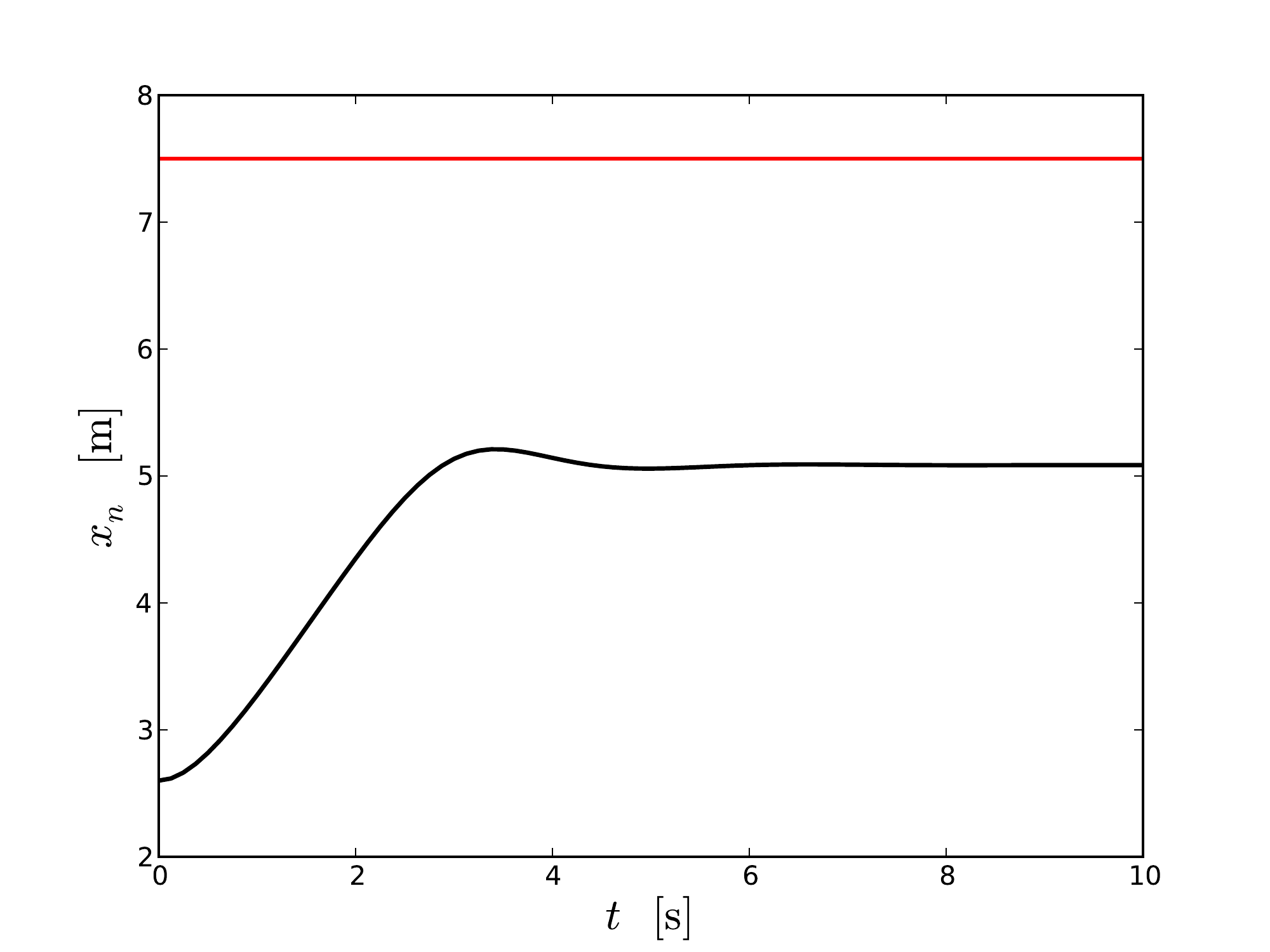}
\caption{Simulation with $a=300$ N and $b=0.5$ m.}
\label{fig2}
\end{figure}

The model  \cite{Lakoba2005} with $a=300$\, N and
$b=0.5$\, m also oscillates. See Fig. \ref{fig2}.

In \cite{Moussaid2009a} an experimental evaluation leads to $a=4.5 \pm 0.3$,\,
$b=1.25$ m (mean value) and $\tau=0.54 \pm 0.05$ s. That means that the system according to Eq. (\ref{eq:cond1}) is oscillating.

%D_{ij} = \lambda (v0,0) + e_{ij}=(1,0)
%||D_{ij}|| = \lambda\cdot v0 + 1
% B = \mu ||D_{ij}|| = \mu  (\lambda\cdot v0 + 1)
% \mu=0.35
% \lambda = 2
% v0=1.29
%B=0.35*(2*1.29 +1)
% B = 1.25
\section{Conclusion}

The presented results show that the social force model has to be extended by using velocity terms in order to produce realistic collision-free behaviors with reasonable values for the parameters. 
% This has notably been done in \cite{Gao2013} or in \cite{Jiang2001,Helbing1998b} within the optimal velocity traffic model \cite{Bando1995}.

In its original form without velocity dependency (reasonable) values of parameters lead inevitably to erroneous oscillating movement.

%\bibliographystyle{abbrv}
%\bibliography{/home/chraibi/litDB/ped}

%\printindex
\end{document}